# Two-dimensional natural hyperbolic materials: From polaritons modulation to applications

Guangyi Jia,[a] Jinxuan Luo,[a] Huaiwen Wang,*[ab] Qiaoyun Ma,[a] Qinggang Liu,[c] Haitao Dai,*[d] Reza Asgari*[e]

Natural hyperbolic materials (HMs) in two dimensions (2D) have an extraordinarily high anisotropy and a hyperbolic dispersion relation. Some of them can even sustain hyperbolic polaritons with great directional propagation and light compression to deeply sub-wavelength scales due to their inherent anisotropy. Herein, the anisotropic optical features of 2D natural HMs are reviewed. Four hyperbolic polaritons (i.e., phonon polaritons, plasmon polaritons, exciton polaritons, and shear polaritons) as well as their generation mechanism are discussed in detail. The natural merits of 2D HMs hold promise for practical quantum photonic applications such as valley quantum interference, mid-infrared polarizer, spontaneous emission enhancement, near-field thermal radiation, and a new generation of optoelectronic components, among others. These analyses' conclusion outlines existing issues and potential interesting directions for 2D natural HMs. These findings could spur more interest in anisotropic 2D atomic crystals in the future, as well as the quick generation of natural HMs for new applications.

## 1 Introduction

Two-dimensional (2D) materials are crystalline solids with relatively very large lateral dimensions as compared to their thickness. Since the discovery of graphene in 2004,[1] 2D materials have become a hot research subject for their rich quantum states and exotic physicochemical characteristics, such as the integer and fractional quantum Hall effects,[2,3] high temperature superconductivity,[4] and numerous topological states of matter,[5,6] etc. In particular, recent advances in the studies of high-quality, atomically thin van der Waals (vdW) crystals have been igniting tremendous enthusiasm for exploring novel 2D natural hyperbolic materials (HMs, also known as indefinite media).[7-11] Natural HMs are a special type of anisotropic media, and they are characterized by effective permittivity tensors where the principal components of the electric or magnetic fields have opposite signs. Therefore, the isofrequency surface (IFS) of transverse-magnetic (TM) polarized waves in HMs takes the form of an open hyperboloid. Compared with the closed IFS of general media, the hyperbolic dispersion theoretically allows for an endlessly large wave vector accompanied by a greatly increased photonic density of states.[12-14] The enhanced light-matter interaction has significant consequences for strong augmentation of spontaneous emission and Cherenkov emission with low energy electrons.[15-18] Moreover, the evanescent waves with large wave vectors in general materials would be converted into propagating waves that propagate without attenuation in this strongly anisotropic medium. Thus, natural HMs can be applied in super-resolution imaging, well below the farfield diffraction limit.[19,20]

At terahertz wavelengths, E. Gerlach et al discovered in 1976 that a bismuth (Bi) crystal after being cooled down to liquid nitrogen temperature has opposite-sign permittivity components along the in-plane and out-of-plane directions.[21] This could be the first time that natural HMs are confirmed in experiment. Subsequently, S. C. Bayliss et al experimentally proved in 1982 that $1T-ZrS_2$ also has different signs of principal elements of permittivity at 77 K in the near ultraviolet region.[22] However, the concept of HMs had not been proposed until D. R. Smith et al found hyperbolic dispersion characteristics in a metamaterial composed of split ring resonators and straight wires in 2003.[23] This kind of metamaterial was later referred to as the hyperbolic metamaterial. It may be due to the rigorous requirement of ultralow temperature and non-tunability of inherent structure of natural HMs, researchers have been for a long time to pay attention on structuring various artificial metastructures to modulate the hyperbolic optical properties.[23-30] Until 2011, J. Sun et al reported a nature indefinite permittivity in vdW crystal of graphite and obtained the hyperbolic isofrequency contour according to ellipsometry data at room temperature (RT).[31] This retriggers the research interest of people in hunting for natural HMs in vdW crystals including graphite analogues, perovskites,

[a.] School of Science, Tianjin University of Commerce, Tianjin 300134, P. R. China. E-mail: wanghw@tjcu.edu.cn
[b.] Tianjin Key Laboratory of Refrigeration Technology, Tianjin University of Commerce, Tianjin 300134, P. R. China
[c.] State Key Laboratory of Precision Measurement Technology and Instruments, Tianjin University, Tianjin 300072, P. R. China.
[d.] Tianjin Key Laboratory of Low Dimensional Materials Physics and Preparing Technology, School of Science, Tianjin University, Tianjin 300072, P. R. China. E-mail: htdai@tju.edu.cn
[e.] School of Physics, Institute for Research in Fundamental Sciences, IPM, Tehran 19395-5531, Iran. E-mail: asgari@ipm.ir

and topological semimetals.[31-39] Especially, natural HMs are endowed with many fantastic merits in comparison with traditional hyperbolic metamaterials.[40-45] For instance, they require no artificial structuring and contain no internal interfaces for the electrons to scatter off. Moreover, because of their homogeneous nature, the hyperbolic dispersion, which is in principle only limited by the atomic periodicity of crystal lattice or effects of non-local dielectric response, is expected to extend much further in reciprocal space.[32]

Up to now, there have been many literatures to review artificial hyperbolic metamaterials and metasurfaces.[7,15,24-26,46] Although some reviews about natural HMs have been also published, they are restricted to either only reviewing hyperbolic phonon polaritons in materials (e.g., hBN, α-MoO$_3$, α-V$_2$O$_5$, and β-Ga$_2$O$_3$) or solely introducing hyperbolic plasmon polaritons in black phosphorus (BP) and WTe$_2$.[7,42-44,47] The comparisons and discussions on phonon polaritons, plasmon polaritons, and exciton polaritons have been also made by W. Ma et al and Y. Wu et al.[48,49] However, their reviews focus on either in-plane anisotropic polaritons or the manipulation strategies for polaritons in vdW materials, in which many recently reported 2D natural HMs (e.g., ZrSiSe, β-Ga$_2$O$_3$, perovskites and tetradymites) along with the related hyperbolic properties are missing.[48,49] In particular, hyperbolic exciton and shear polaritons have been experimentally identified in tetradymites and β-Ga$_2$O$_3$ in recent.[50,51] A comprehensive comparison of hyperbolic phonon, plasmon, exciton, and shear polaritons remains lacking. Moreover, various novel and improved applications for natural HMs (e.g., valley quantum interference and high temperature polarizer[45,52]) have been also presented during the last two years. Thereby, it is meaningful to provide a thorough overview of these four types of hyperbolic polaritons, as well as the most recent developments in 2D natural HMs' applications.

This paper examines 2D materials which have been shown to sustain natural hyperbolicity in experiments. The basic electromagnetic response and hyperbolic wavelengths of 2D natural HMs are discussed first. The advanced methods for modifying hyperbolic response are then introduced, followed by four hyperbolic polaritons found in natural HMs and their fascinating nanophotonic features and generation mechanism. Following that, we look at a few examples of applications that make use of natural hyperbolicity in 2D materials. In the end, we discuss the difficulties encountered and the prospects for 2D natural HMs in the future.

## 2 Basic physics of HMs

The concept of hyperbolic dispersion stems from anisotropic crystals with the tensors of relative permittivity or permeability

$$\hat{\varepsilon} = \begin{pmatrix} \varepsilon_{xx} & 0 & 0 \\ 0 & \varepsilon_{yy} & 0 \\ 0 & 0 & \varepsilon_{zz} \end{pmatrix} \text{ or } \hat{\mu} = \begin{pmatrix} \mu_{xx} & 0 & 0 \\ 0 & \mu_{yy} & 0 \\ 0 & 0 & \mu_{zz} \end{pmatrix} \quad (1)$$

where the subscripts $xx$, $yy$, and $zz$ indicate the components along principal axes of $x$-, $y$-, and $z$-directions, respectively. Herein, we concentrate on electric anisotropic materials and assume permeability components are equal in all directions ($\mu_{xx} = \mu_{yy} = \mu_{zz} = 1.0$). If one or more principal components of $\hat{\varepsilon}$ has different signs, the crystal becomes either a uniaxial or a biaxial anisotropic material. We posit that the optical axis is along the $z$-direction, then the electric uniaxial and biaxial materials have $\varepsilon_{xx} = \varepsilon_{yy} \neq \varepsilon_{zz}$ and $\varepsilon_{xx} \neq \varepsilon_{yy} \neq \varepsilon_{zz}$, respectively.

The electric and magnetic components of optical plane wave can be described as

$$\boldsymbol{E} = \boldsymbol{E}_0 e^{i(i\omega t - \boldsymbol{k} \cdot \boldsymbol{r})} \quad (2)$$

$$\boldsymbol{H} = \boldsymbol{H}_0 e^{i(i\omega t - \boldsymbol{k} \cdot \boldsymbol{r})} \quad (3)$$

where $\boldsymbol{E}_0$ and $\boldsymbol{H}_0$ are the magnitudes of electric and magnetic fields, respectively, $k = 2\pi/\lambda$ is the wave vector and $\lambda$ is the wavelength. The parameters $\boldsymbol{r}$, $\omega$ and $t$ indicate position, angular frequency and time, respectively. The plane wave expressions at eqns (2) and (3) are inserted into the following Maxwell's equations

$$\nabla \times \boldsymbol{E} = -\mu_0 \frac{\partial \boldsymbol{H}}{\partial t} \quad (4)$$

$$\nabla \times \boldsymbol{H} = \varepsilon_0 \hat{\varepsilon} \frac{\partial \boldsymbol{E}}{\partial t} \quad (5)$$

where $\varepsilon_0$ and $\mu_0$ are electric permittivity and magnetic permeability in free space, respectively. Then, two equations are obtained as

$$\boldsymbol{k} \times \boldsymbol{E} = \omega \mu_0 \boldsymbol{H} \quad (6)$$

$$\boldsymbol{k} \times \boldsymbol{H} = -\omega \varepsilon_0 \hat{\varepsilon} \boldsymbol{E} \quad (7)$$

According to eqns (6) and (7), one can get an eigenvalue equation with[46]

$$\boldsymbol{k} \times (\boldsymbol{k} \times \boldsymbol{E}) + \omega^2 \mu_0 \varepsilon_0 \hat{\varepsilon} \boldsymbol{E} = 0 \quad (8)$$

which can be expanded in matrix form as

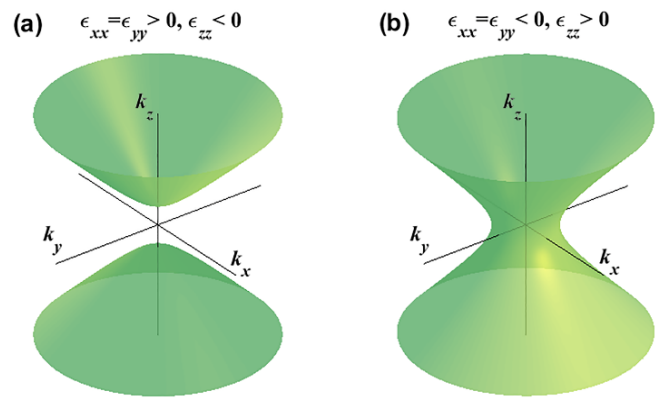

**Fig. 1** IFSs of TM waves in (a) Type-I and (b) Type-II uniaxial HMs

$$\begin{pmatrix} k_0^2\varepsilon_{xx} - k_y^2 - k_z^2 & k_xk_y & k_xk_z \\ k_xk_y & k_0^2\varepsilon_{yy} - k_x^2 - k_z^2 & k_yk_z \\ k_xk_z & k_yk_z & k_0^2\varepsilon_{zz} - k_x^2 - k_y^2 \end{pmatrix}\begin{pmatrix} E_x \\ E_y \\ E_z \end{pmatrix} = 0 \quad (9)$$

where $k_0$ is the magnitude of wave vector in vacuum.

The transverse-electric (TE) and TM polarized waves can coexist in a HM. For a uniaxial medium, nontrivial solutions to eqn (9) yield the dispersion relation as

$$\left(k_x^2 + k_y^2 + k_z^2 - \varepsilon_{xx}k_0^2\right)\left(\frac{k_x^2 + k_y^2}{\varepsilon_{zz}} + \frac{k_z^2}{\varepsilon_{xx}} - k_0^2\right) = 0 \quad (10)$$

According to eqn (10), the IFS of TE wave takes the form of a sphere. However, the IFS of TM wave is an unbounded hyperboloid that can support high-$k$ waves if the signs of two components of $\hat{\varepsilon}$ are opposite ($\varepsilon_{xx}\cdot\varepsilon_{zz} < 0$).[15,46] Furthermore, uniaxial HMs can be classified into two types. Type-I materials with $\varepsilon_{xx}= \varepsilon_{yy} > 0$ and $\varepsilon_{zz} < 0$ have a predominantly dielectric nature and their IFS is two sheeted, as shown in Fig. 1(a). By contrast, Type-II materials with $\varepsilon_{xx}= \varepsilon_{yy} < 0$ and $\varepsilon_{zz} > 0$ have more metallic properties thus are highly reflective.[7,15] Fig. 1(b) shows that the IFS of TM wave in a Type-II material is single sheeted. A general medium can be achieved when all the components of $\hat{\varepsilon}$ have the same sign, and the IFS of TM wave becomes closed and forms a sphere or ellipsoid.

For biaxial materials, the hyperbolic dispersion relations can be obtained through eqn (9) with three conditions: $k_x = 0$, $k_y = 0$ and $k_z = 0$. Fig. 2 shows the IFSs of two kinds of biaxial HMs. The first kind has only one negative permittivity component (e.g., $\varepsilon_{xx} > 0$, $\varepsilon_{yy} > 0$, $\varepsilon_{zz} < 0$, $\varepsilon_{xx} \neq \varepsilon_{yy}$) and its IFS is a two-fold hyperboloid [Fig. 2(a-c)]. The second one has two negative permittivity components (e.g., $\varepsilon_{xx} < 0$, $\varepsilon_{yy} < 0$, $\varepsilon_{zz} > 0$, $\varepsilon_{xx} \neq \varepsilon_{yy}$) and its IFS is a one-sheet hyperboloid [Fig. 2(d-f)]. According to the Reststrahlen band (RB) being defined between the transverse and longitudinal optical phonon frequencies,[38,40] the biaxial hyperbolic characteristics can be further generalized to three types of materials which are associated with three optical axes. Fig. 3 shows the real parts of permittivity tensors of α-MoO$_3$ and α-V$_2$O$_5$ crystals, which are derived by using a Lorentz model.[38,40] We can see that there are three RBs in the plotted spectral region (in colour) where at least one of the permittivity components is negative. RB1 and RB2 originate from the in-plane phonons along the $y$- and $x$-directions, respectively. RB3 results from the out-of-plane phonon along the $z$-direction.

Theoretically, more than 40 vdW crystals have been predicted to be natural HMs.[32,33,53,54] Nevertheless, without any modifications, only a few of them have been experimentally demonstrated to harbor natural hyperbolic properties.[31-40,55,56] Table 1 and Fig. 4 summarize the experimentally confirmed hyperbolic wavelength ranges of 16 vdW crystals which are without any artificial modifications. It is seen that the tetradymites BiSe, $Bi_2Se_3$ and $Bi_2Te_3$ support broadband hyperbolic response in the visible frequency range. Because all of the hyperbolic phonon polaritons, plasmon polaritons, and exciton polaritons are experimentally observed in graphite analogues,[35-38,40,50] graphite-like vdW crystals are received more attention than perovskites in the field of natural HMs. Unfortunately, natural hyperbolic response has not yet been discovered in experiment for graphite-like vdW materials in the wavelength ranges of 1306−6250, 7300−9651, and 23420−53700 nm. Very recently, layered nodal-line semimetal ZrSiSe (features a non-symmorphic P4/nmm space group, and the weak vdW interaction between adjacent Se-Zr-Si-Zr-Se quintuple layers provides a natural cleaving surface)[57] has also been measured to possess Type-II hyperbolic regime at the telecommunications frequencies.[39] In addition to the natural HMs listed in Table 1, the opposite sign of permittivity components has been also experimentally measured in some bulk optical crystals, such as that in quartz,[58] sapphire,[59] calcite,[60,61] and $YVO_4$.[61,62] Because these bulk optical crystals are not vdW media and are difficult to be exfoliated into 2D quantum materials, they are not the focus of this review.

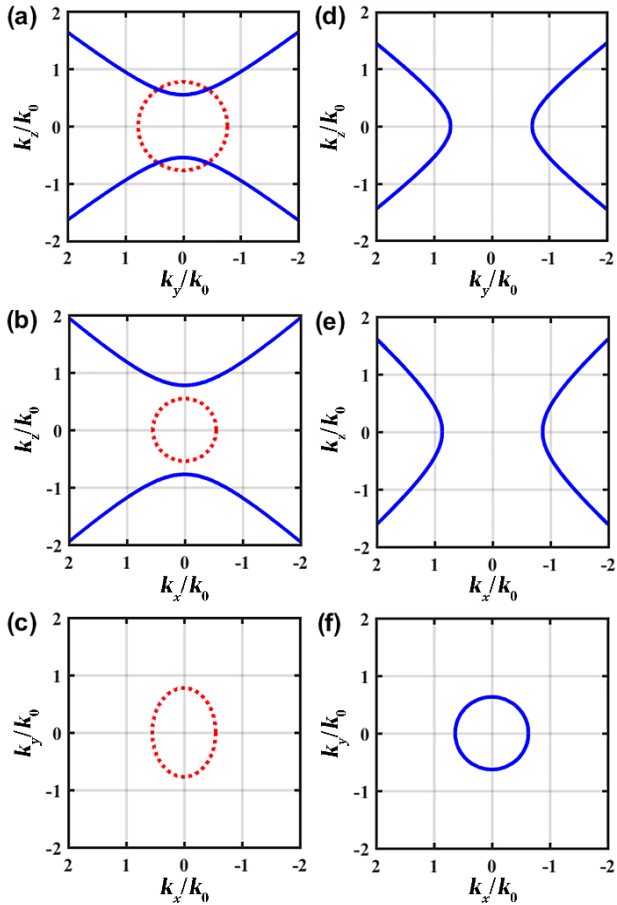

**Fig. 2** IFSs of biaxial HMs with permittivity components of (a-c) $\varepsilon_{xx}$ = 0.6, $\varepsilon_{yy}$ = 0.3, $\varepsilon_{zz}$ = -0.4 and (d-f) $\varepsilon_{xx}$ = -0.6, $\varepsilon_{yy}$ = -0.3, $\varepsilon_{zz}$ = 0.4. (a)(d), (b)(e), and (c)(f) correspond to the views from $k_x$-, $k_y$-, and $k_z$-axes, respectively. Blue solid and red dot lines are from TM and TE waves, respectively.

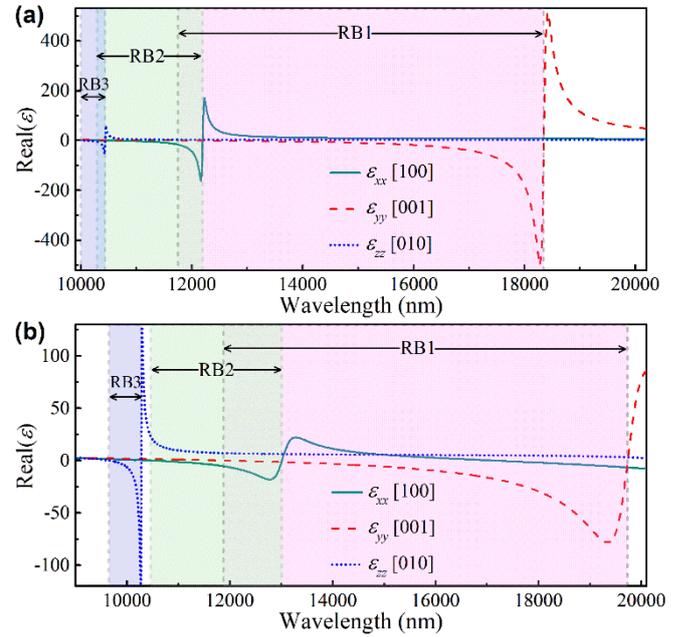

**Fig. 3** Real parts of the permittivities of (a) α-MoO$_3$ and (b) α-V$_2$O$_5$ crystals along the principal $x$ ([100]), $y$ ([001]) and $z$ ([010]) axes (green, red, and blue lines, respectively). The Reststrahlen bands RB1, RB2 and RB3 are shaded in pink, green, and blue, respectively.

Table 1 Overview of experimentally confirmed natural hyperbolic vdW crystals and their hyperbolic wavelength ranges

| Materials | | Hyperbolic spectral range (nm) | | | | | Temperature | Refs. |
|---|---|---|---|---|---|---|---|---|
| | | Uniaxial crystals | | Biaxial crystals | | | | |
| | | Type-I | Type-II | RB1 | RB2 | RB3 | | |
| | Graphite | | 207–282 | | | | RT | 31, 32 |
| | 1T-ZrS$_2$ | | 354–~391 | | | | 77 K | 22 |
| | MgB$_2$ | | 333.10–477.38 | | | | RT | 33 |
| | Bi | 53700–63200 | | | | | 5 K, 80 K | 21, 55 |
| | BiSe | 525–710 | 210–265 | | | | RT | 34 |
| Graphite-like crystals | Bi$_2$Se$_3$ | 500–1040 | 210–230 | | | | RT | 35 |
| | Bi$_2$Se$_3$ | 708.57–1180.95 | 288.37–688.89 | | | | RT | 35 |
| | Bi$_2$Te$_3$ | 1033.33–1305.26 | 288.37–885.71 | | | | RT | 36 |
| | hBN | 12195.12–13157.89 | 6250.00–7299.27 | | | | RT | |
| | T$_d$-WTe$_2$ | | | 16051.36–23419.20 | | | below 200 K | 37 |
| | α-V$_2$O$_5$ | | | 11875.86–19735.13 | 10463.15–13014.11 | 9650.24–10282.38 | RT | 38 |
| | α-MoO$_3$ | | | 11750.88–18348.62 | 10288.07–12195.12 | 9900.99–10438.41 | RT | 40 |
| | Sr$_2$RuO$_4$ | 920.17–15614.19 | | | | | | |
| | Sr$_3$Ru$_2$O$_7$ | 1019.70–13626.93 | | | | | RT | 33 |
| Perovskites | (BA)$_2$PbI$_4$ | around 513 | | | | | | |
| | (BA)$_2$(MA)Pb$_2$I$_7$ | around 571 | | | | | 80 K – RT | 56 |
| Nodal-line semimetal | ZrSiSe | ~1300 – ~1700 | | | | | RT | 39 |

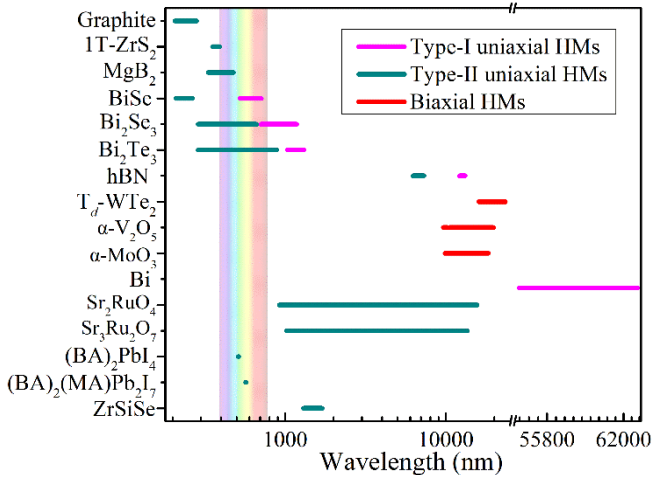

**Fig. 4** Hyperbolic wavelength ranges of experimentally demonstrated natural hyperbolic vdW crystals. The iridescence bar indicates the visible-light regime.

## 3 Hyperbolic polaritons in 2D natural HMs

### 3.1 Hyperbolic phonon polaritons

Hyperbolic polaritons, which are described as light-matter hybrid quasiparticles, arise from the coupling of photons to elementary excitation (e.g., plasmons, phonons, excitons, and shears) in natural HMs. They can be utilized to control and manipulate the electromagnetic waves at the deep subwavelength scale, being essential for constructing compact nanophotonic devices and circuits.[7,43,51] Among four polaritons, hyperbolic phonon polaritons were first observed in 2D hBN by D. N. Basov et al via using infrared nanoimaging in 2014.[63]

Subsequently, vdW crystals of $\alpha$-$MoO_3$[64-67] and $\alpha$-$V_2O_5$[38] were also demonstrated to be natural HMs sustaining hyperbolic phonon polaritons in the mid-infrared to terahertz region.

Optical phonons originate from coherent atomic lattice vibrations, in which the displacements of two atoms happens in opposite directions and the vibrations are at the frequencies of 10–1500 $cm^{-1}$, in crystalline media.[43] Optical phonons have transverse and longitudinal modes. The former (latter) corresponds to lattice vibrations perpendicular to (along) the direction of wave propagation at the frequency $\omega_T$ ($\omega_L$). In nonpolar crystals, transverse and longitudinal optical phonons are degenerate and the vibrations are charge neutral. Therefore, the permittivity of nonpolar crystal is little affected by nonpolar phonons. By contrast, optical phonons in polar crystals give rise to charge separation with shifting longitudinal optical phonon to a higher frequency, i.e., $\omega_L > \omega_T$. Importantly, the transverse optical phonon possesses a net polarizability, which can enable strong interaction with light (strong absorption) and lead to coherent oscillations of ionic charges opposite to the incident electromagnetic field. As a consequence, the real part of one permittivity component could become negative at the RB frequency range between $\omega_L$ and $\omega_T$.

hBN is a typical representative of the optical phonon based systems. Fig. 5(a) presents that hBN gives two separate RBs. The phonon polaritons possess Type-I (Type-II) hyperbolic dispersion in lower (upper) RB 760–820 $cm^{-1}$ (1370–1600 $cm^{-1}$).[36] Among various characterization techniques, scattering-type scanning near-field optical microscopy (s-SNOM) has been demonstrated to be a powerful method to launch and image hyperbolic phonon polaritons with a resolution deeply below the diffraction limit.[43,68,69] The s-SNOM can probe both the near-field amplitude and phase of phonon polaritons via real-space nanoimages.[43,68,69] Based on the s-SNOM, J. Duan et al proved that the polariton wavelength $\lambda_p$ (505.8 nm) of hBN deposited on $SiO_2$ is larger than that (236.0 nm) of hBN on Au film [see Fig. 5(b-c)].[36] S. Dai et al further demonstrated that the $\lambda_p$ in the suspended hBN is elongated by ~12% [$\Delta \lambda_p = (\lambda_{p,sus} - \lambda_{p,sup})/\lambda_{p,sup}$] in comparison with that in the supported hBN on $SiO_2$ [see Fig. 5(d)].[70] The discrepancies in polariton wavelengths are attributed to the change of substrate permittivity, indicating that engineering the permittivity environment is a potential method to modulate the properties of hyperbolic phonon polaritons. Besides, the elimination of substrate loss (i.e., sample suspension) can also effectively reduce the damping and elongate the propagation length of phonon polaritons in hBN [cf. red and blue line profiles in Fig. 5(d)].[36,70]

In addition to modifying the permittivity environment with rigid substrates, various vdW heterostructures consisting of hBN and phase-change materials or the other 2D materials have been also designed to alter the phonon polaritons.[71-73] Even so, the previous works demonstrate a rigid phase ≈π/4 without evident deviations for the polariton reflection. Until 2022, M. Chen et al realized the alteration of polariton reflection phase for hyperbolic surface polaritons in hBN.[74] As shown in the inset in Fig. 5(e), two types of phonon polaritons are imaged in the hBN microstructures at $\omega$ = 1419 $cm^{-1}$ by using s-SNOM. Parallel to the edges, linear fringes were imaged with the strongest

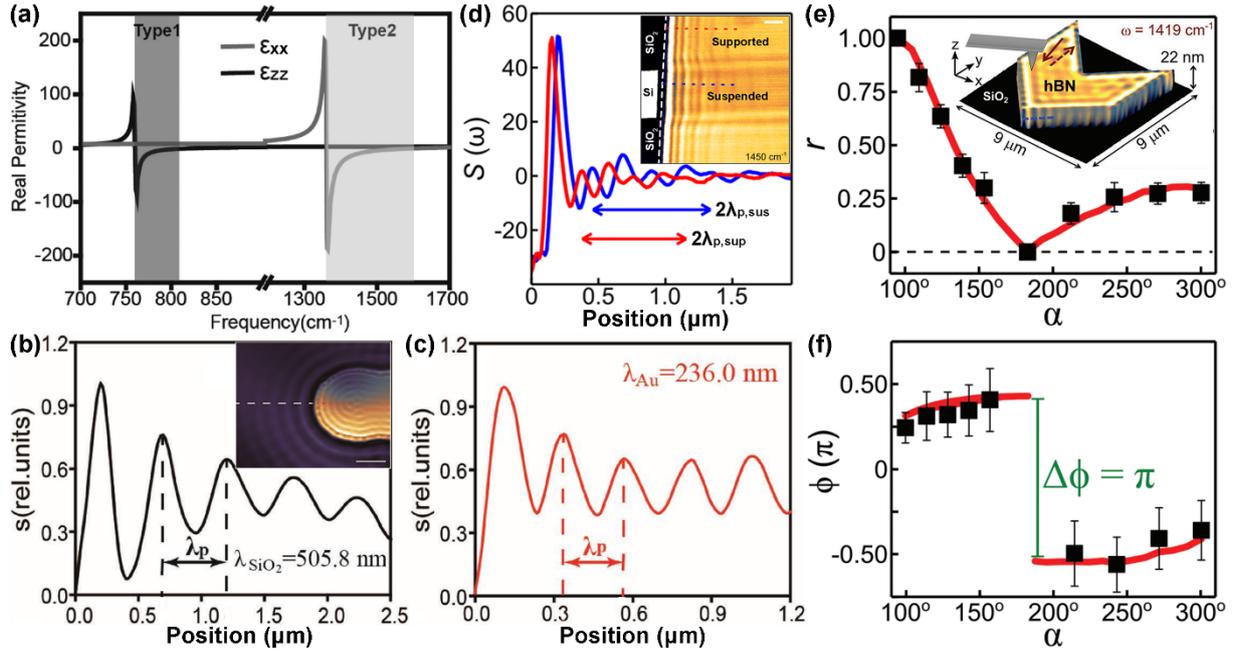

**Fig. 5** (a) Real components of the permittivity tensor for hBN crystal. The corresponding spectral regions of Type-I and Type-II are shaded in grey. (b), (c) s-SNOM line profiles along the dashed white (on SiO$_2$ substrate) and red (on gold substrate) lines in the inset in (b), respectively. The wavelength of propagating hyperbolic phonon polaritons is compressed about twice on gold substrate. (a-c) Reproduced with permission from ref. 36. Copyright 2017, Wiley-VCH. (d) s-SNOM line profiles taken along the dashed cuts in the s-SNOM image (inset) of suspended and supported hBN on SiO$_2$ at $\omega$ = 1450 cm$^{-1}$. Reproduced with permission from ref. 70. Copyright 2019, American Chemical Society. The evolution of (e) amplitude and (f) phase of the polariton reflection coefficient with the corner angle $\alpha$ of hBN microstructure. The inset in (e) shows the schematic of the s-SNOM experiment with a representative s-SNOM image on the hBN microstructure. (e-f) Reproduced with permission from ref. 74. Copyright 2022, Wiley-VCH.

oscillation close to the edge, followed by weakly damped ones into the interior. Fig. 5(e-f) shows the evolution of amplitude and phase of the polariton reflection coefficient with the corner angle $\alpha$ of hBN microstructure. The reflection amplitude [Fig. 5(e)] first decreases from 1 to 0 as $\alpha$ is raised from 90° to 180°, then increases from 0 to 0.3 as $\alpha$ changes from 180° to 300°. On the other hand, the reflection phase [Fig. 5(f)] of hyperbolic surface polaritons has a three-step evolution and even reveals a $\pi$ jump by varying the corner angle $\alpha$.

Despite the natural hyperbolicity, hBN is uniaxial such that it does not carry anisotropic responses at the interface, i.e, the in-plane isofrequency curve is circular. To excite in-plane hyperbolic phonon polaritons, recently emerged biaxial polar media of α-V$_2$O$_5$ and α-MoO$_3$ provide new degrees of freedom to control photons in 2D owing to their distinctive in-plane anisotropies.[38,40,67,75,76] For example, to modify the spectral range of RB, the intercalation of alkali-metal atoms into vdW crystal α-V$_2$O$_5$ was proposed by J. Taboada-Gutiérrez et al.[38] Comparing with the pristine α-V$_2$O$_5$, Na-intercalated α-V$_2$O$_5$ (forming the crystal α'-(Na)V$_2$O$_5$) exhibits a RB shift of ~30 cm$^{-1}$ (60% of the initial RB width), as shown in Fig. 6(a). But the similar lifetimes for phonon polaritons, as depicted in Fig. 6(b-e), evidence that the intercalation of external atoms into α-V$_2$O$_5$ does not substantially affect the polaritonic properties of the crystal.[38] Besides, the polariton isofrequency contours can be precisely manipulated via harnessing the relative twist angles in bilayers of HM, e.g., α-MoO$_3$ as illustrated in Fig. 6(f-g), in which tunable topological transitions from open (hyperbolic) to closed (elliptical) dispersion contours can be achieved, even extreme states such as low-loss tunable polariton canalization can occur at certain critical twist angles.[75-77] Furthermore, the in-plane hyperbolic phonon polaritons in α-MoO$_3$ can be focused by a gold plasmonic nanoantenna with delicately designed convex extremity, as shown in Fig. 6(h).[67] One can also tailor the lateral sizes (diameters) of focus spots of phonon polaritons by either controlling the illumination frequency or the curvature radius of the antenna extremity. Especially, a focal spot as small as 340 nm in diameter [see Fig. 6(i)], being 1/32 of the free-space wavelength, can be launched at the illumination frequency of 926 cm$^{-1}$ and positive curvature radius of 2.5 μm.[67]

As for the sample suspension, a wavelength elongation $\Delta\lambda_p$ = 60% is demonstrated by S. Dai et al in 2022 in the suspended α-MoO$_3$ with a positive phase velocity [see Fig. 7(a)], which is ~5 times larger than that reported in hBN.[78] This kind of more significant wavelength elongation indicates that hyperbolic phonon polaritons in α-MoO$_3$ have a smaller confinement than those in hBN. S. Dai et al also examined hyperbolic phonon polaritons in α-MoO$_3$ with a negative phase velocity. In contrast to $\Delta\lambda_p > 0$ in the lower band in Fig. 7(a), the polariton wavelength in the upper band gets shortened $\Delta\lambda_p$ = -38% ($\lambda_{p,sus}$ = 539 nm and

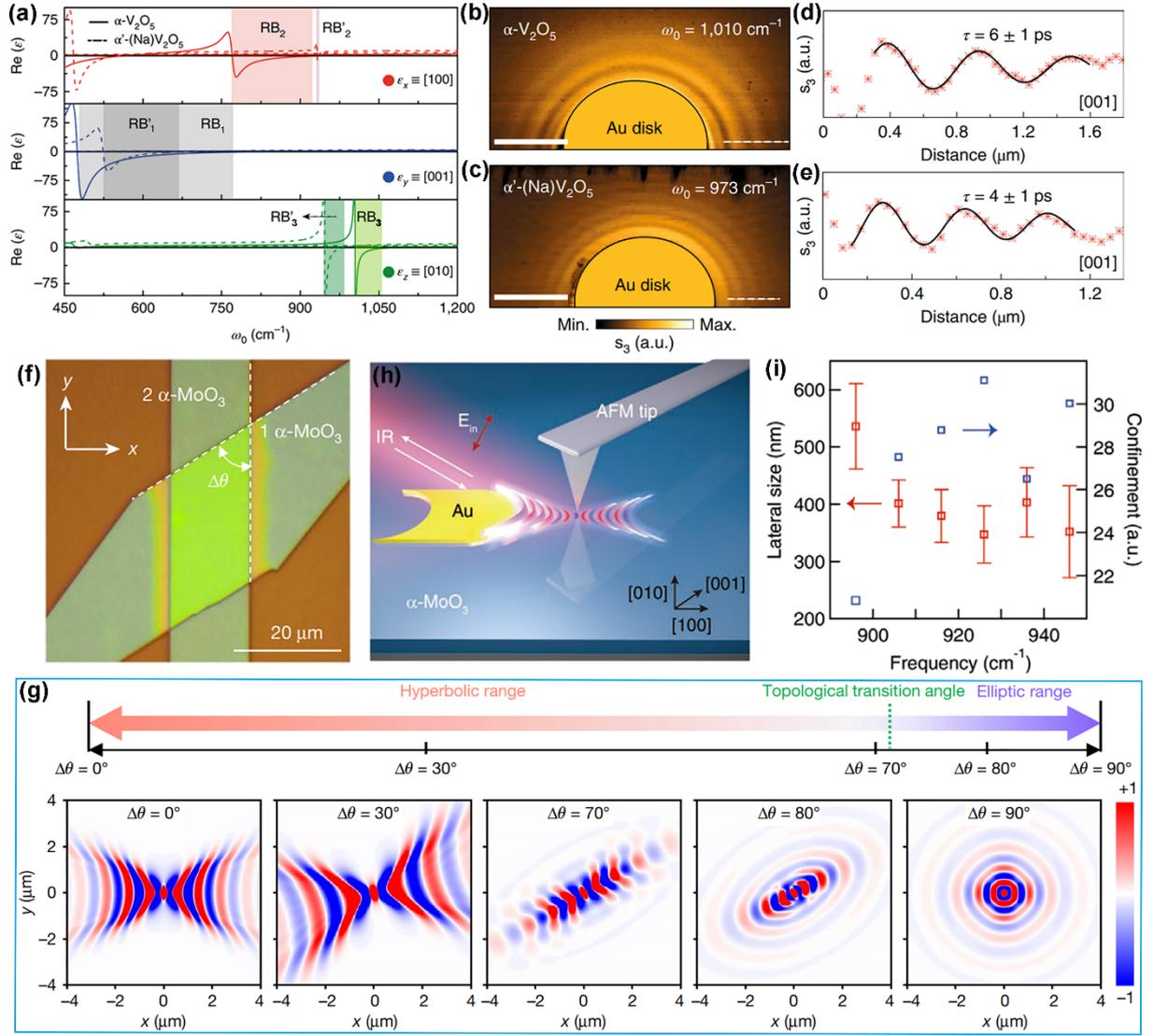

**Fig. 6** (a) Real part of the permittivities for α-V$_2$O$_5$ (continuous lines) and α'-(Na)V$_2$O$_5$ (dashed lines) extracted from ab initio calculations along three principal axes. Near-field amplitude images of (b) α-V$_2$O$_5$ and (c) α'-(Na)V$_2$O$_5$ flakes. s-SNOM line traces along the [001] direction of (d) α-V$_2$O$_5$ and (e) α'-(Na)V$_2$O$_5$ flakes indicated by white dashed lines in (b) and (c). (a-e) Reproduced with permission from ref. 38. Copyright 2020, Springer Nature. (f) Optical image of a twisted bilayer α-MoO$_3$ flake with Δϑ = −57°. The white dashed lines denote the [001] crystal direction of the two layers. (g) Topological nature of the polariton dispersion and corresponding field distributions (real part of z-component of the electric field) at different twist angles. (f-g) Reproduced with permission from ref. 75. Copyright 2020, Springer Nature. (h) Launching in-plane hyperbolic phonon polaritons in α-MoO$_3$ with delicately designed curved gold plasmonic nanoantennas. (i) Lateral sizes of the focus spots of phonon polaritons and the corresponding electromagnetic confinement factor $\lambda_0/\lambda_p$ ($\lambda_0$ and $\lambda_p$ indicate the illumination wavelength and polariton wavelength, respectively) at different frequencies. (h-i) Reproduced with permission from ref. 67. Copyright 2022, Wiley-VCH.

$\lambda_{p,sup}$ = 875 nm) by sample suspension, as shown in Fig 7(b). The opposite alteration of polariton wavelengths in lower and upper RBs has also been reported in gradually suspended α-MoO$_3$.[79] As shown in Fig. 7(c), a nanoscale gradient air gap is formed between the SiO$_2$ substrate and the top α-MoO$_3$ flake. The gap size $t_{gap}$, i.e., the separation between the lower surface of the suspended flake and substrate, gradually decreases from hundreds of nanometers to zero. Fig. 7(d-e) reveals that the dependence of $\lambda_p$ on $t_{gap}$ in lower and upper RBs contradict each other. Specifically, as the $t_{gap}$ increases to 205 nm [Fig. 7(d)], the tuning range of wavelength elongation $\Delta\lambda_p$ can reach to 160% for phonon polaritons at ω = 937 cm$^{-1}$ [$\lambda_0$ = 10672.36 nm is in RB2 in Fig. 3(a)]. By contrast, the compression of polariton wavelength is observed for phonon polaritons in RB3 in the suspended α-MoO$_3$ flake. As illustrated in Fig. 7(e), the compression of $\lambda_p$ ranges from 12% to 36% with increasing the gap size from 28 to 185 nm at ω = 990 cm$^{-1}$ (i.e., $\lambda_0$ = 10101.01 nm), resulting in greatly enhanced electromagnetic field localizations. The opposite variation tendency of polariton wavelengths is also accompanied with contrary change of

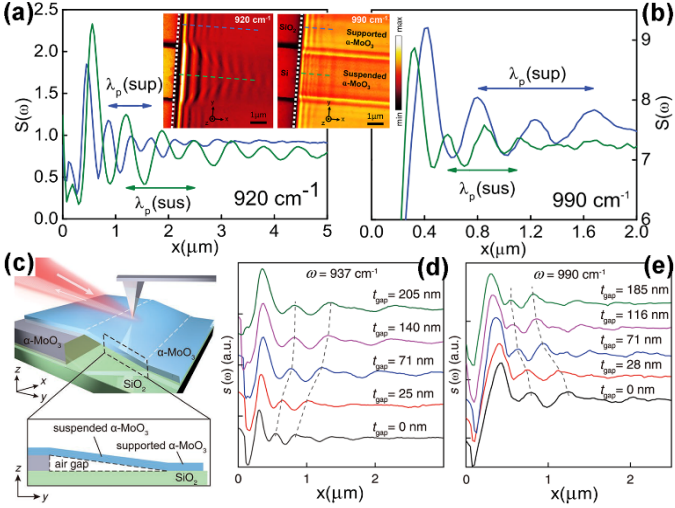

**Fig. 7** (a) s-SNOM line profiles taken along dashed lines in the left inset from suspended and supported α-MoO$_3$ at $\omega$ = 920 cm$^{-1}$. (b) s-SNOM line profiles taken along dashed lines in the right inset from suspended and supported α-MoO$_3$ at $\omega$ = 990 cm$^{-1}$. The insets show the s-SNOM amplitude images of hyperbolic phonon polaritons at $\omega$ = 920 and = 990 cm$^{-1}$. (a-b) Reproduced with permission from ref. 78. Copyright 2022, AIP Publishing. (c) Schematic showing the gradually suspended α-MoO$_3$ flake with a gradient air gap. (d) and (e) show the variation tendencies of s-SNOM line profiles with respect to the gap size $t_{gap}$ at $\omega$ = 937 and 990 cm$^{-1}$, respectively. Grey dashed lines in (d) and (e) are guide for the eye tracking separations between adjacent maxima. (c-e) Reproduced with permission from ref. 79. Copyright 2022, Wiley-VCH.

confinement ratio and propagation length of polaritons in lower and upper RBs of α-MoO$_3$, as shown in Table 2. These features strongly suggest a much wider tuning range for hyperbolic phonon polaritons in α-MoO$_3$ than those reported in hBN.

## 3.2 Hyperbolic plasmon polaritons

In 2012, Z. Fei et al and J. Chen et al detected the propagating surface plasmons in graphene by using s-SNOM with infrared excitation light.[80,81] By virtue of exclusive 2D plasmonic properties, e.g., strong energy-confinement capability,[80,81] distinct carrier-density dependence,[82] and strong coupling to optical phonons of substrate,[83] etc., the discovery of graphene plasmons stimulates growing efforts to explore more and novel plasmonic 2D materials. In particular, the experimental observation of hyperbolic phonon polaritons has further stirred research interest in search of hyperbolic plasmon polaritons in anisotropic 2D materials. In 2016, A. Nemilentsau et al predicted the natural in-plane hyperbolic plasmons in 2D BP, demonstrating that hyperbolic plasmons are closely related with the coupling interplay between intraband and interband transitions.[84]

More specifically, the complex optical conductivity (or permittivity) of 2D crystal is composed of anisotropic intraband transitions (i.e., the Drude response) and interband transitions along the principal axes of $x$- and $y$-directions. The Drude component contributes to the positive imaginary part of optical conductivity at the low-energy side while the interband transition component leads to the negative values at the high-energy side, as shown in Fig. 8(a). In view of the crystal asymmetry, the imaginary parts of optical conductivities $\sigma_{xx}$ and $\sigma_{yy}$ can change their signs at different energies, giving rise to an energy interval for hyperbolic plasmons where Im($\sigma_{xx}$)Im($\sigma_{yy}$) < 0. In this hyperbolic case, the plasmon dispersion in $q$ space is a hyperbola [green and red lines in Fig. 8(b)], resulting in directionally propagated polaritons in Fig. 8(c-d). In contrast, the

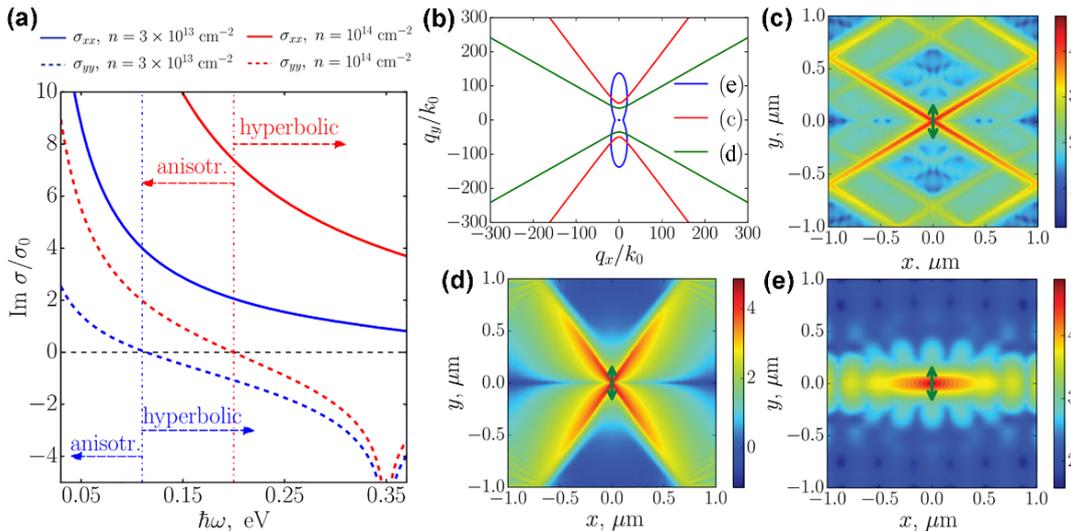

**Fig. 8** (a) Imaginary part of the optical conductivity for 2D BP along two principal axes of $x$- and $y$-directions at different carrier densities $n$. (b) Isofrequency contours for plasmons in panels (c-e). (c-e) The electric field distributions in real space (log scale) for surface plasmons excited by a $y$-polarized electric dipole [green arrow in panels (c-e)]. (c) $n$ = 10$^{14}$ cm$^{-2}$, $\hbar\omega$ = 0.3 eV (d) $n$ = 3×10$^{13}$ cm$^{-2}$, $\hbar\omega$ = 0.3 eV. (e) $n$ = 10$^{14}$ cm$^{-2}$, $\hbar\omega$ = 0.165 eV. The corresponding $q$-space dispersions are shown in (b). Reproduced with permission from ref. 84. Copyright 2016, American Physical Society.

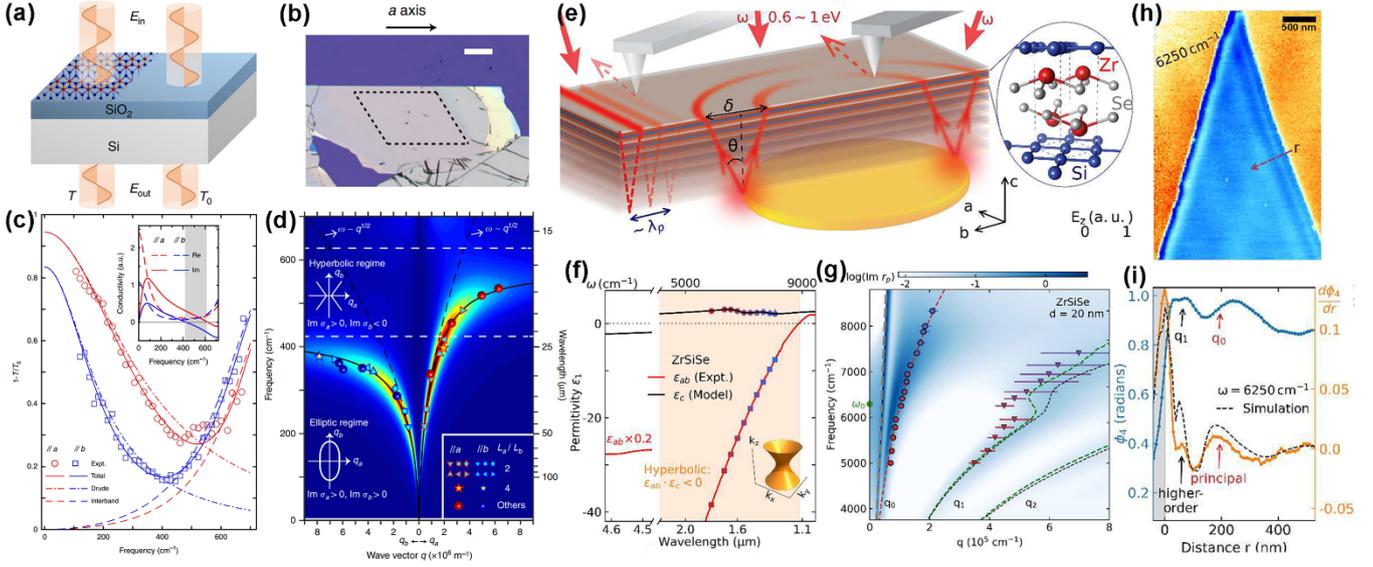

**Fig. 9** (a) Schematic of the setup for measuring extinction spectra and (b) the optical microscope image of exfoliated WTe$_2$ thin film. (c) Far infrared absorption spectra of exfoliated WTe$_2$ thin film in (b) and the derived optical conductivities (inset) along two in-plane axes at 20 K. The shaded area in the inset represents the hyperbolic frequency regime. (d) Plasmon dispersion for polarizations along $a$ and $b$ axes measured in WTe$_2$ rectangle arrays. The calculated loss function is shown as a pseudo color map. Solid and dashed black curves stand for the fitted dispersion with and without considering the interband transitions, respectively. Dashed horizontal white lines denote the hyperbolic regime obtained from the loss function calculation. (a-d) Reproduced with permission from ref. 37. Copyright 2020, Springer Nature. (e) Schematic sketch of the experimental setup for excitation and detection of hyperbolic plasmon polaritons in ZrSiSe. Inset shows the layered lattice structure of ZrSiSe. (f) In-plane permittivity ($\varepsilon_{ab}$, red line) obtained from far-field optical measurements. The $\varepsilon_{ab}$ values at selected frequencies (squares) are used to extract $\varepsilon_c$ (circles). The black line is a Drude-Lorentz fit of the experimental out-of-plane permittivity data. Inset is the IFS in momentum space inside the hyperbolic regime ($\varepsilon_{ab} < 0$, $\varepsilon_c > 0$). (g) Dispersion relations of the hyperbolic plasmon polaritons for ZrSiSe film at a thickness of 20 nm. (h) Near-field phase image of a 20 nm thick crystal of ZrSiSe at $\omega = 6250$ cm$^{-1}$. (i) Line profile of the phase contrast (blue dotted line) along the dashed line in (h). Orange dotted line corresponds to the derivative of the phase line profile, revealing features at multiple spatial periodicities. Dashed line is the simulation of the derivative profile with two periodicities, corresponding to the principal (red arrow) and higher-order (black arrow) hyperbolic plasmon polaritons. (e-i) Reproduced with permission from ref. 39. Copyright 2022, CC BY 4.0 License.

isofrequency contour of plasmon for the purely anisotropic regime is an ellipse [the blue line in Fig. 8(b)], and the plasmon propagates along one of the crystallographic axes of BP [the $x$-axis in this case, as shown in Fig. 8(e)].

Because of the highly anisotropic in-plane dynamics, theoretical researches on plasmon polaritons in BP, e.g., engineering 2D BP plasmons via electrostatic bias and straining, have been reported by many groups.[44,85-87] However, in the experimental aspects, there are still many difficulties to realize hyperbolic plasmon polaritons in BP, including the low carrier density, the large bandgap, and the susceptibility of BP to oxygen and humidity. The first two makes the plasmon wave vector in the hyperbolic regime extremely large, adding difficulties in detecting the anisotropic plasmon polaritons by both the farfield and nearfield methods.[44,88] The last one makes BP quite unstable and easily deteriorated in air. Although the carrier density and bandgap can be tuned by the methods of straining, doping and gating, defects and impurities will be introduced into BP by these methods as well, which can create more damping channels for plasmons. Therefore, researchers shift their attention to the other layered anisotropic materials.

In 2020, H. Yan et al firstly realized the measurement of natural hyperbolic plasmon polaritons in exfoliated Weyl semimetal WTe$_2$ at ultralow temperatures.[37] Fig. 9(a-b) shows the schematic of a setup for measuring extinction spectra and the optical microscope image of an exfoliated WTe$_2$ thin film. Fig. 9(c) presents the far infrared absorption spectra of WTe$_2$ and the derived optical conductivities along the principal $x$- and $y$-axes (corresponding to $a$ and $b$ axes in Fig. 9, respectively). One can clearly find a frequency interval of 427–623 cm$^{-1}$ where Im($\sigma_{xx}$) > 0 and Im($\sigma_{yy}$) < 0 in the inset in Fig. 9(c). In order to investigate the plasmon dispersion, the authors fabricated a set of rectangle arrays along the two optical axes of WTe$_2$ films. The calculated loss function (pseudo color map) and the measured plasmon frequencies for polarizations along two principal axes are displayed in Fig. 9(d). The dashed black lines denote standard 2D plasmon dispersion with $\omega \propto q^{1/2}$. It is seen that the plasmon peaks follow the $q^{1/2}$ scaling at low energies. As the wave vector $q$ increases, the dispersions along both in-plane axes gradually deviate from the purely free carrier case due to the coupling to interband transitions, and approach an energy limit around 400 cm$^{-1}$ (600 cm$^{-1}$) for plasmons along the $y$- ($x$-) axis at large wave vectors. The hyperbolic regime locates at the range of 429–632 cm$^{-1}$, being well in line with the regime obtained through the optical conductivity in the inset in Fig. 9(c).

In 2022, Y. Shao et al demonstrated that the nodal-line semimetal ZrSiSe hosts propagating hyperbolic plasmon polaritons across the entire telecommunication wavelength range of 5900–7700 cm$^{-1}$, and visualized both the principal and higher-order modes of hyperbolic plasmon polaritons by using the s-SNOM.[39] As illustrated in Fig. 9(e), the near-infrared laser illuminates layered ZrSiSe residing on a circular gold antenna. The antenna launches the principal mode of hyperbolic plasmon polaritons which travel along conical trajectories (red arrows) inside the sample with an angle $\theta$ from the surface normal. The hyperbolic rays reaching the top surface of the sample are separated by a distance $\delta$ and are scattered back to the detector by the sharp tip, producing two characteristic rings in raster scanned images. Besides the antenna, the sharp sample edge [see Fig. 9(h) and the left of sample in Fig. 9(e)] can also launch polaritons, which contributes to characteristic higher-order hyperbolic modes in Fig. 9(i). The dispersion relations of hyperbolic plasmon polaritons, which are calculated for a 20 nm thick crystal of ZrSiSe residing on Si/SiO$_2$ substrate by using experimental permittivities in Fig. 9(f), are shown in Fig. 9(g). Multiple dispersive branches develop in the hyperbolic frequency range, corresponding to the propagated principal and higher-order modes within the bulk ZrSiSe. The calculated hyperbolic dispersions agree with the experimental momenta (colored circles and triangles), unequivocally corroborating the appearance of hyperbolic plasmon polaritons in ZrSiSe.

In addition to 2D BP, WTe$_2$ and ZrSiSe, some the other layered anisotropic crystals, e.g., $\beta$-allotrope of carbon phosphide[89] and massive tilted 2D Dirac materials,[54,90] have been also theoretically predicted to support hyperbolic plasmons. Nevertheless, the experimental detection of hyperbolic plasmon polaritons has not yet been realized in these materials due to the demand for strong coupling between intraband and interband transitions.

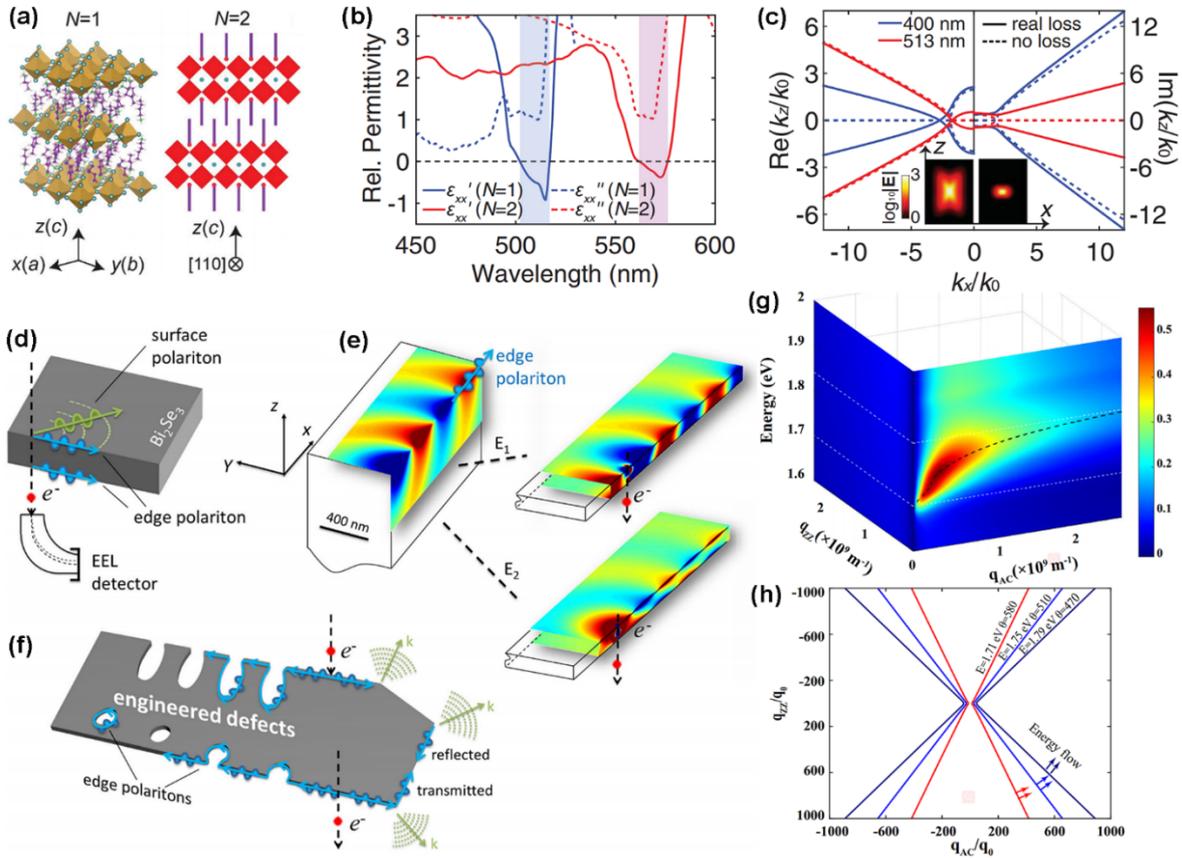

**Fig. 10** (a) Schematic of the crystal structures of (BA)$_2$PbI$_4$ ($N$ = 1) and (BA)$_2$(MA)Pb$_2$I$_7$ ($N$ = 2). (b) In-plane $\varepsilon'$ and $\varepsilon''$ for $N$ = 1 and $N$ = 2. The colored areas highlight the spectral regions of $\varepsilon'$ < 0 (blue for $N$ = 1 and red for $N$ = 2). (c) Isofrequency curves for $N$ = 1 at 513 nm (red) and 400 nm (blue) with real loss ($\varepsilon''$ taken to be the measured values; solid lines) and no loss ($\varepsilon''$ taken to be 0; dashed lines). Inset: simulated electric field amplitude at 513 nm (left) and 400 nm (right) with real loss. Width and height of the plotted domain are 50 nm and 75 nm, respectively. (a-c) Reproduced with permission from ref. 56. Copyright 2018, American Physical Society. (d) Schematic diagrams of the surface and edge polaritons in Bi$_2$Se$_3$ and the corresponding energy-dispersive detection system. (e) $z$-component of the electric field associated with edge polaritons which propagate along the edges of a Bi$_2$Se$_3$ large cube at the energy of 4 eV. Red and blue colors stand for positive and negative values, respectively. (f) The interaction of hyperbolic edge polaritons at corners and with precisely engineered grooves and nanocavities. (d-f) Reproduced with permission from ref. 50. Copyright 2021, Springer Nature. (g) and (h) are the calculated loss function and the in-plane dispersions of hyperbolic exciton polaritons in monolayer BP at different frequencies, respectively. (g-h) Reproduced with permission from ref. 93. Copyright 2021, Springer Nature.

## 3.3 Hyperbolic exciton polaritons

The aforementioned hyperbolic phonon and plasmon polaritons generally appear in mid-infrared or terahertz frequency region. For the hyperbolicity in the visible-light regime, the perovskites and tetradymites (see Fig. 4 and Table 1) should be visited, which support exciton polaritons. Excitons are excited electron-hole pairs bounded by Coulomb interactions in semiconductors, forming the exciton polaritons when coupled to electromagnetic waves. Even if exciton polaritons have been measured in some transition metal dichalcogenides, e.g., $WS_2$ and $WSe_2$, the utilized materials are isotropic such that it is unable to produce hyperbolic exciton polaritons.[91,92]

In 2018, P. Guo et al firstly reported out-of-plane propagating exciton polaritons with hyperbolic dispersions in 2D perovskite crystals.[56] Fig. 10(a) depicts the crystal structures of 2D perovskites which comprise chemical compositions of $(BA)_2(MA)_{N-1}Pb_NI_{3N+1}$ with $N$ = 1, 2, BA = $CH_3(CH_2)_3NH_3^+$, and MA = $CH_3NH_3^+$. Fig. 10(b) displays the real ($\varepsilon'$) and imaginary part ($\varepsilon''$) of in-plane relative permittivities for $N = 1$ and $N = 2$. One can find that the hyperbolic regimes for $N = 1$ and $N = 2$ are centered at 513 and 571 nm, respectively. Fig. 10(c) presents the TM isofrequency curves for $N = 1$, demonstrating the transformation of IFS from an ellipsoidal shape at 400 nm to a hyperboloid at 513 nm. The photonic density of states is proportional to the area of IFS at a given wavelength, thus the photonic density of states is greatly amplified near the exciton resonance in comparison with that at 400 nm. The enhanced photonic density of states can be further validated by simulating the near fields. As shown in the inset in Fig. 10(c), the electric field strength at 513 nm is much stronger than that at 400 nm.

In 2021, R. Lingstädt et al experimentally demonstrated that $Bi_2Se_3$ nanoplatelets with specifically shaped nanocavities supports hyperbolic edge exciton polaritons (HEEPs) at RT.[50] As illustrated in Fig. 10(d), the specimen area of interest is either scanned by a focused electron probe or irradiated with a parallel beam of fast electrons. Both surface polaritons and edge polaritons are excited during the inelastic interaction of fast electrons with specimen. HEEPs propagate along the edges of $Bi_2Se_3$ nanoplatelets and their mode volumes are sharply confined to the edges [Fig. 10(e)]. Besides, the HEEPs, which bound to the upper and lower edges of $Bi_2Se_3$ nanoplatelets, hybridize into a symmetric and an anti-symmetric mode, respectively. The launched optical modes are partially reflected at corners, partially guided around them to the adjacent side planes, or couple to far-field radiation, as shown in Fig. 10(f).

Besides experimentally investigated perovskites and $Bi_2Se_3$, monolayer BP has been recently predicted to hold in-plane hyperbolic exciton polaritons from 1.703–1.844 eV.[93] Fig. 10(g) plots the loss function $-Im(1/\varepsilon)$ as a pseudo color map, where $\varepsilon(\boldsymbol{q}, \omega)$ is the dynamical permittivity along two distinctive directions of BP, $\boldsymbol{q}$ and $\omega$ indicate the wave vector and frequency, respectively. Then the maxima of $-Im(1/\varepsilon)$ in the map are extracted as the isofrequency contour in $\boldsymbol{q}$-space, and all of them present hyperbolic shapes, as depicted in Fig. 10(h). The hyperbolic exciton polaritons in BP are attributed to the in-plane anisotropic excitonic behaviors which produce an energy window for the sign-changing of imaginary part of optical conductivity along the armchair direction.

## 3.4 Hyperbolic shear polaritons

The lattice symmetry also plays a critical role in dictating the physical properties of a crystal material. Thus far, the permittivity tensors of 2D natural HMs being used for studying phonon polaritons, plasmon polaritons, and exciton polaritons are mainly limited to being diagonalized. Reducing the crystal symmetry can provide emergent opportunities to control light propagation, polarization and phase.[38,40,65,75,94-96] In 2022, N. C. Passler et al observed a new polariton class, i.e., hyperbolic shear polaritons, in a low-symmetry monoclinic crystal $\beta$-$Ga_2O_3$ in which the permittivity tensor is naturally non-diagonalizable at infrared frequencies.[51]

The description of the dielectric response of $\beta$-$Ga_2O_3$ requires inclusion of identical off-diagonal elements in the monoclinic plane with

$$\overline{\overline{\varepsilon(\omega)}} = \begin{bmatrix} \varepsilon_{xx}(\omega) & \varepsilon_{xy}(\omega) & 0 \\ \varepsilon_{xy}(\omega) & \varepsilon_{yy}(\omega) & 0 \\ 0 & 0 & \varepsilon_{zz}(\omega) \end{bmatrix} \quad (11)$$

The frequency-dependent spectra for permittivity tensor elements are shown in Fig. 11(a). The monoclinic crystal structure as well as the coordinate systems used to define the dielectric response of $\beta$-$Ga_2O_3$ is sketched in Fig. 11(b). The sample surface of $\beta$-$Ga_2O_3$ is the monoclinic (010) plane ($x$-$y$ plane). It is seen that $\beta$-$Ga_2O_3$ has low-symmetry Bravais lattices and non-orthogonal principal crystal axes. The permittivity tensor has major polarizability directions that strongly depend on the frequency and exhibits shear terms analogous to viscous flow.[97]

These features stem from the non-trivial relative orientation (neither parallel nor orthogonal) of several optical transitions that contribute to a net polarization (cannot be aligned with the crystal axes) at a given frequency. In turn, this property leads to an exotic light propagation which is not supported by higher-symmetry crystals.[98,99,100] It should be noted that $\beta$-$Ga_2O_3$ is not a vdW material, but single-crystal $\beta$-$Ga_2O_3$ can be separated into quasi-2D layers of $\beta$-$Ga_2O_3$—potentially—even in monolayers—along its facile cleavage planes [(100) and (001)].[101,102] Strictly speaking, hyperbolic shear polaritons are accepted as one kind of phonon polaritons. To highlight the key influence of non-diagonalizable dielectric permittivity in optical properties of low-symmetry crystals, hyperbolic shear polaritons are separately classified in this work.

For lossless case, the solutions for shear polariton wave vectors at two different frequencies of 713 and 718 cm$^{-1}$ are given in Fig. 11(c). Two open hyperboloid IFSs can be observed, in particular, both the wave vector magnitude and the propagation direction of polaritons strongly disperse with the frequency. When we take into account the natural material loss resulting from inherent phonon-scattering processes, the rotation of hyperbolic shear polaritons with respect to the coordinate system of the monoclinic plane can be seen even at an individual

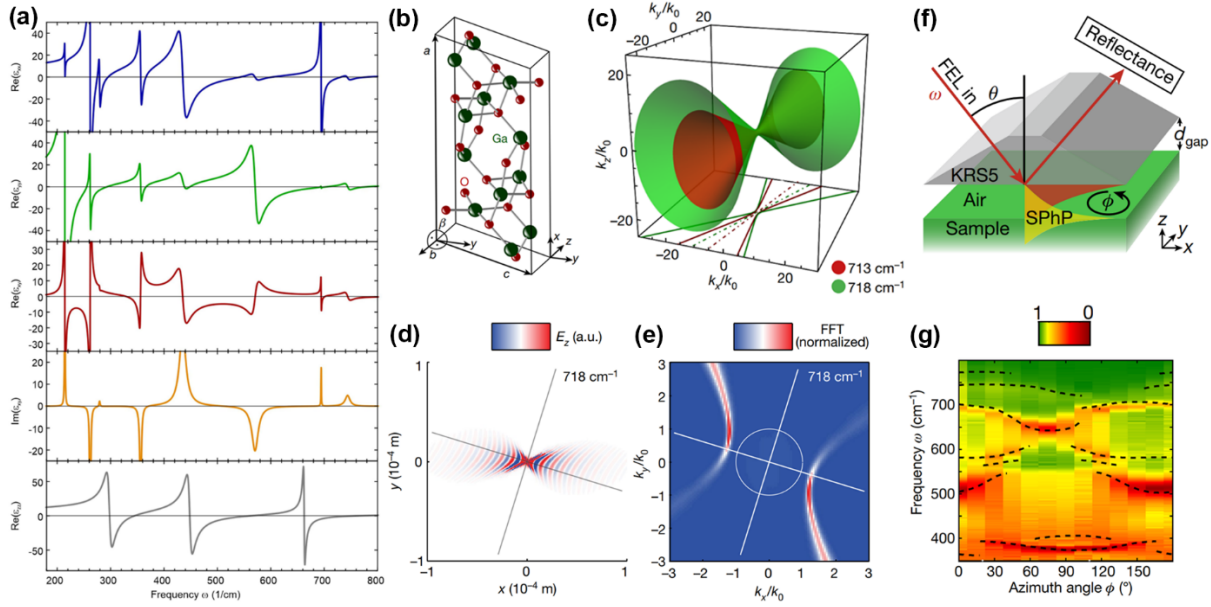

**Fig. 11** (a) Permittivity tensor elements $\varepsilon_{xx}$, $\varepsilon_{yy}$, $\varepsilon_{xy}$ (=$\varepsilon_{yx}$) and $\varepsilon_{zz}$ of charge carrier-free β-Ga$_2$O$_3$ at infrared frequencies. (b) The monoclinic crystal structure of β-Ga$_2$O$_3$ and (c) the IFSs at two different frequencies (red and green). In (c), the contour lines at $k_z = 0$ and their mirror axes are plotted as solid and dash-dotted lines at the bottom, respectively. (d) Real-space electric field and the corresponding (e) 2D Fourier transformation at the x-y plane surface of β-Ga$_2$O$_3$. (f) Otto-type prism-coupling experimental setup for the observation of surface phonon polaritons on β-Ga$_2$O$_3$ sample. (g) Reflectance map: experimental azimuth and frequency dependence of hyperbolic shear polaritons on β-Ga$_2$O$_3$. (a-g) Reproduced with permission from ref. 51. Copyright 2022, Springer Nature.

frequency, as illustrated by the isofrequency contours in Fig. 11(d). Moreover, the 2D Fourier transformation of real-space electric field profile [see Fig. 11(e)] indicates that the wavefronts of shear polariton propagation are tilted with respect to the major propagation direction, with no apparent mirror symmetry. Such tilted wavefronts are one of the most notable and unique characteristics of hyperbolic shear polaritons, which is a direct consequence of the low symmetry of the crystal material.

The impacts of crystal asymmetry on polariton propagation are further verified by using an Otto-type prism-coupling experimental setup [Fig. 11(f)]. The experimental azimuthal dispersion of surface phonon polaritons on monoclinic β-Ga$_2$O$_3$ exhibits no mirror symmetry, as shown in Fig. 11(g). This asymmetric polaritonic dispersion at rotating the azimuth angle is a direct consequence of the lack of symmetry in the vibrational structure of β-Ga$_2$O$_3$.[98] Thus, the propagation direction of hyperbolic shear polaritons shows a continuous rotation as the frequency is varied, accordingly, hyperbolic shear polaritons are not propagated along fixed axes.

On the basis of the above discussions, Fig. 12 sketches out a diagram for these four types of polaritons. Surface plasmon polaritons, exciton polaritons, and phonon polaritons arise from the coupling of electromagnetic waves with collective oscillations of free charges, Coulomb-bound electron-hole pairs (i.e, excitons), and phonon resonances, respectively. The shear polaritons emerge from the coupling of infrared light to phonons in monoclinic crystals. In contrast to traditional phonon polaritons observed in crystals with a symmetric structure, four of six sides are rectangular but two are tilted parallelograms in monoclinic crystals [see Fig. 11(b)]. Because of this structure of distorted cuboid, shear waves not only run very directed across the crystal surface but are also no longer mirror-symmetric. Polaritons can be considered as a hybrid of light-matter oscillations. The matter oscillation component gives rise to negative permittivity of the polaritonic material, leading to optical-field confinement at the interface with a positive-permittivity environment (e.g., the air). The polaritons exhibit strong confinement $\chi$, as defined by the ratio of incident light wavelength $\lambda_0$ to polariton wavelength $\lambda_p$. Additionally, one of the figures of merits to characterize polariton loss is the propagation length $L_p$. The propagation length defines the distance that a polariton mode will propagate until the amplitude of electric field decays to $1/e$.

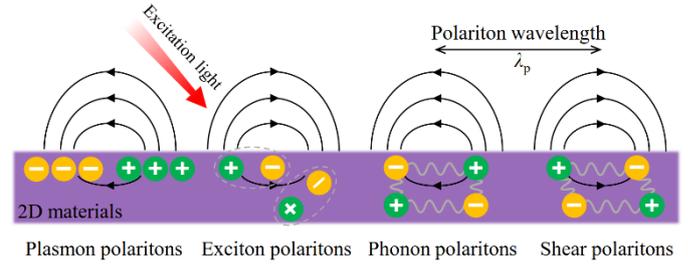

**Fig. 12** Schematic illustration of four types of polaritons in 2D natural HMs.

Table 2 summarizes some characteristics of four hyperbolic polaritons in natural HMs. One can see that the confinement

Table 2 Real-space characteristics of hyperbolic polaritons in natural HMs

| Polaritons | Materials | $d$ (nm) | $\lambda_0$ (nm) | $\chi$ | $L_p$ (nm) | Features | Refs. |
|---|---|---|---|---|---|---|---|
| Phonon polaritons | hBN | 77 | 6535.95 | 27.69 | NG | Gold launch, in-plane mode | 36 |
| | | | | 12.92 | NG | $SiO_2$ spacer thickness 285 nm, in-plane mode | |
| | hBN | NG | 6896.55 | ~16.08* | 487.21* | $SiO_2$ spacer thickness 285 nm, in-plane mode | 70 |
| | | | | ~14.25* | 633.90* | Suspension height 285 nm, in-plane mode | |
| | α-MoO$_3$ | 247 | 10869.57 | 13.59 | 848 | Supported by Si/SiO$_2$ substrate, [100] direction | 78 |
| | | | | 8.49 | 2037 | Suspended on an etched trench in SiO$_2$ spacer, [100] direction | |
| | | | 10101.01 | 11.54 | 1160 | Supported by Si/SiO$_2$ substrate, [100] direction | |
| | | | | 18.74 | 953 | Suspended on an etched trench in SiO$_2$ spacer, [100] direction | |
| | α-V$_2$O$_5$ | 130 | 9900.99 | ~17.9* | 1400 | Gold launch, in-plane ellipticity, [001] direction | 38 |
| | α′-(Na)V$_2$O$_5$ | 107 | 10277.49 | ~28.5* | 1150 | Gold launch, in-plane ellipticity, [001] direction | 38 |
| Plasmon polaritons | WTe$_2$ | ~100 | 50000.00 | ~6 | NG | Diamond substrate, nanoribbon width 240 nm, $T = 10$ K | 37 |
| | | | 19531.25 | ~40 | NG | | |
| | ZrSiSe | 20 | 1400 | 3.08* | NG | Supported by Si/SiO$_2$ substrate, [001] direction | 39 |
| | | | 1600 | 2.69* | NG | | |
| HEEPs | Bi$_2$Se$_3$ | 60 | 310 | ~0.89 | NG | Launched by a finite grating structure consisting of round-shaped grooves at an edge of Bi$_2$Se$_3$ | 50 |
| Shear polaritons | β-Ga$_2$O$_3$ | Bulk | 13927.58 | NG | 1420* | Launched by an Otto-type prism-coupling configuration at an incident angle of 28° | 51 |
| | | | 14025.25 | NG | 1700* | | |

NG means "not given"; *Data extracted or calculated from related graphs in references.

ratios $\chi$ of hyperbolic phonon polaritons are much larger than 1.0, indicating the polariton wavelength $\lambda_p \ll \lambda_0$. HEEPs are launched by the grating structure with round-shaped grooves in Fig. 10(f), the effective HEEP wavelength is about half of the perimeter (~700 nm) of each groove structure.[50] As for the propagation lengths of experimentally observed hyperbolic plasmon and exciton polaritons, their exact values might be lacking in the literature. The propagation length is greatly affected by damping pathways which are tightly correlated with polaritonic materials themselves and surrounding dielectric environment.[49] It is seen from Table 2 that the $L_p$ values for hyperbolic phonon polaritons can changes from ~500 to ~2000 nm through switching polaritonic materials and tuning the external physical conditions.

# 4 Applications of 2D natural HMs

The natural hyperbolicity along with the nanophotonic advantages endows 2D HMs with a series of potential technological applications such as optical sensing,[103,104] nanostructure diagnosis,[105,106] photonic metasurfaces,[107,108] solar energy harvesting,[109] subdiffraction focusing and imaging,[110,111] etc. These applications have been reviewed by some groups.[7,43,44,47] To avoid repetition, several recently proposed and improved promising directions are exemplified as follows.

## 4.1 Valley quantum interference

In 2D gapped Dirac systems, e.g., transition metal dichalcogenides and biased bilayer graphene,[112-114] the electronic band structure is composed of two degenerate valleys $K$ and $K'$ in $k$ space, in which the optical selection rule is sensitive to the polarization of exciting photons. Valley coherence can be quantified through the degree of linear polarization $P = (I_x − I_y)/(I_x + I_y)$ of the emission. Here, $I_x$ and $I_y$ indicate the intensities of two linearly polarized emissions. In the presence of a weak pump, the parameter $P$ is nearly equal to the quantum interference $Q = (\gamma_x − \gamma_y)/(\gamma_x + \gamma_y)$ where $\gamma_x$ and $\gamma_y$ imply the Purcell factors for $x$ and $y$ polarized dipoles, respectively. The valley quantum interference is proportional to the anisotropy between the Purcell factors for orthogonal dipoles.

In 2022, A. Bapat et al proposed to measure the valley quantum interference via placing a gapped Dirac system at a certain distance away from the heterostructure consisting of monolayer graphene and hyperbolic crystal of α-MoO$_3$ on a silicon substrate.[45] The geometrical configuration for the

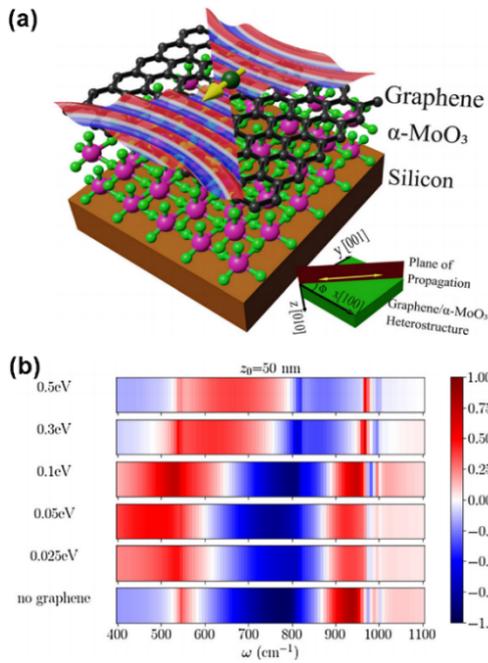

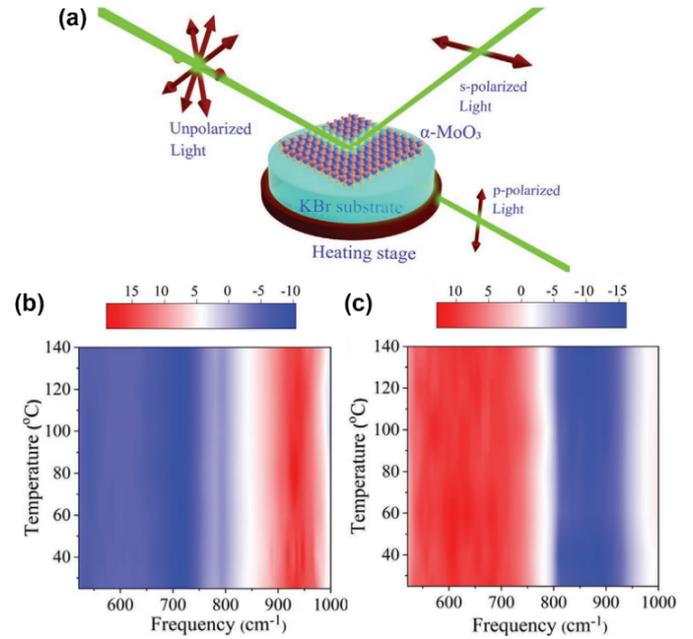

**Fig. 13** (a) The heterostructure of graphene/α-MoO$_3$ deposited upon silicon substrate and the hybrid plasmon-phonon polariton modes propagation at the surface. Inset illustrates the crystal orientation of α-MoO$_3$ and the angle between plane of propagation of light and principal $x$ ([100]) axis of α-MoO$_3$. (b) The quantum interference as a function of frequency at different chemical potentials of graphene. The distance of electric dipole from the surface of the heterostructure is taken to be $z_0$ = 50 nm. The thickness of α-MoO$_3$ film is 50 nm. Reproduced with permission from ref. 45. Copyright 2022, De Gruyter.

**Fig. 14** (a) A device of reflection and transmission type polarizer based on α-MoO$_3$ thin film. The extinction ratios for (b) reflection and (c) transmission type polarizers as a function of temperature and frequency. Reproduced with permission from ref. 52. Copyright 2022, Wiley-VCH.

heterostructure is sketched in Fig. 13(a). The hyperbolic phonon polaritons of α-MoO$_3$ can hybridize with surface plasmon polaritons of graphene to form hybrid plasmon-phonon polaritons at the surface. The gapped Dirac system is placed away from the heterostructure at a distance of $z_0$. The exciton of gapped Dirac system is represented by a dipole. The quantum interference as a function of the frequency is shown in Fig. 13(b). It is seen that the quantum interference is suppressed at graphene chemical potentials of 0.025 and 0.05 eV. As the chemical potential increases to 0.1 eV, 0.3 eV and 0.5 eV, we begin to see large magnitude of quantum interference in RB1 and RB2 spectral regions of α-MoO$_3$, which is induced by the coupling between the emitter and the hybrid plasmon-phonon polaritons.

### 4.2 Mid-infrared polarizer

Intrinsic hyperbolic anisotropy of vdW crystals has been also explored for the application in polarizer and polarization rotator in the mid-infrared spectral region.[115-128] Nonetheless, the earlier studies are mainly limited to applications at RT. In 2022, N. R. Sahoo et al firstly proposed the transmission and reflection type polarizer based on α-MoO$_3$ above RT.[52] As illustrated in Fig. 14(a), the single-crystal α-MoO$_3$ film, which is synthesized via using physical vapor deposition, is transferred on top of a potassium bromide (KBr) window. Due to the in-plane hyperbolic anisotropy of α-MoO$_3$ along [100] and [001] crystallographic directions in the RBs, α-MoO$_3$ reflects the light with s polarization state while transmitting the light with p polarization state. The extinction ratios for the reflection and transmission type polarizers as a function of temperature and frequency are shown in Fig. 14(b-c). It is found that the polarizer device retains high extinction ratios with peak value exceeding 10 dB with a temperature tolerance up to 140 °C. This work provides a lithography-free pathway for the applications of vdW natural hyperbolic crystals in mid-infrared polarizers and quantum cascade devices at high temperatures.[119-123]

### 4.3 Spontaneous emission enhancement

As an intrinsic radiation property of quantum emitter, the spontaneous emission decay rate can be dramatically modified when the quantum emitter locates near a solid matter rather than in vacuum. Furthermore, the superradiance or subradiance could be produced by controlling the relative positions between quantum emitters. Some configurations based on natural HMs have been designed to adjust the interaction between quantum emitters. Nonetheless, most of them are limited to reflection configurations where the quantum emitters are placed on the same side of the structure.[124-128] In 2021, H.-Q. Mu et al investigated the interaction between quantum emitters in the transmission configuration through a sandwich structure composed of graphene and hBN [see the inset in Fig. 15(a)].[129] Fig. 15(a) and (b) show the spontaneous emission decay rate of a single quantum emitter and the interaction between quantum emitters as a function of transition frequency, respectively. In Fig. 15(a), the Purcell factors present two peaks at the hyperbolic

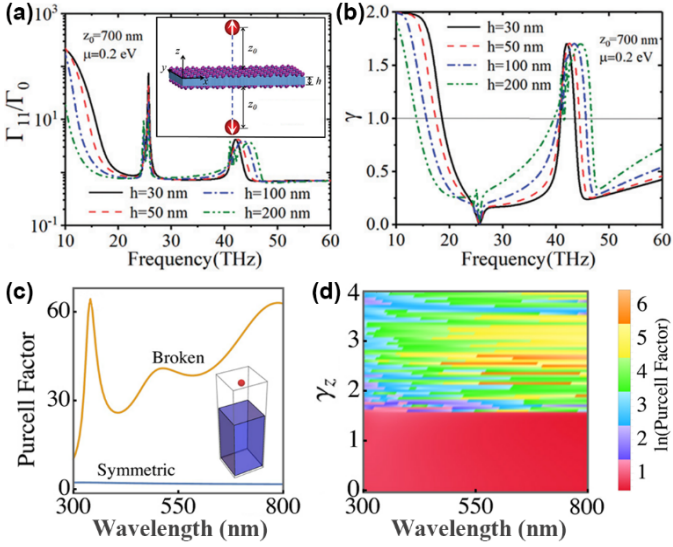

**Fig. 15** (a) Purcell factor as a function of transition frequency of a single quantum emitter for different hBN thicknesses. The inset illustrates the interaction between two quantum emitters through a graphene/hBN/graphene sandwich structure. (b) Interaction between two quantum emitters through the sandwich structure for different hBN thicknesses. (a-b) Reproduced with permission from ref. 129. Copyright 2021, IOP Publishing. (c) Isotropic Purcell factors for $CP$-symmetric [$\gamma$ = (0, 0, 1)] and $CP$-broken [$\gamma$ = (0, 0, 2)] crystals. Here, $\gamma$ = diag($\gamma_x$, $\gamma_y$, $\gamma_z$) represents a real diagonal chiral term. The inset presents the setup in which the cuboid stands for the $CP$-symmetry-related crystal and the red sphere is a dipolar source. (d) Variation of the Purcell factor with the component $\gamma_z$ and the wavelength. The Purcell factor is in a logarithmic scale. The sharp boundary around $\gamma_z \approx 1.55$ manifests the $CP$-broken critical point, at which a large leap of the Purcell factor is observed. (c-d) Reproduced with permission from ref. 130. Copyright 2021, American Physical Society.

bands of type-I (22.8–24.6 THz) and type-II (41.1–48.0 THz), implying that the natural HMs are indeed conducive to enhance the spontaneous emission of single quantum emitter being placed near them. In Fig. 15(b), the interaction between two quantum emitters is very weak such that the system is in the subradiant state when the transition frequency is in type-I hyperbolic band. In marked contrast, one strong peak can be found when the transition frequency is in type-II hyperbolic band, thus the system is in the superradiant state.

Besides, J. Hou et al recently demonstrate a chirality-driven transition from charge-parity ($CP$) symmetric to $CP$-broken phases, which provides a new approach to achieve lossless hyperbolic materials.[130] Based on the COMSOL Multiphysics simulations, they show that the hyperbolic dispersion in the $CP$-broken regime can remarkably enhance the spontaneous emissions of an electric dipole source being placed at 10 nm above the hyperbolic sample, as depicted in Fig. 15(c-d). Unfortunately, the utilized model sample is a bulk quartz crystal. As a type-II Weyl semimetal, previous works have proved that the $CP$ symmetry of $T_d$-WTe$_2$ can be broken via breaking time reversal symmetry,[96,131] which may offer new directions for exploring $CP$-breaking induced natural hyperbolicity in 2D HMs and the related spontaneous emission enhancement.

### 4.4 Near-field thermal radiation

Near-field thermal radiation (NFTR) is a high-efficiency energy transport at the nanoscale. It can be enhanced by the strong interference and coupling effects of evanescent waves, especially when hyperbolic plasmons and/or phonon polaritons are excited.[132-136] Since it can be far ahead of the blackbody limit by several orders of magnitude, plenty of methods have been recently proposed to manipulate the radiative heat flux caused by near-field effects.[134-136]

In 2021, J. Zhang et al investigated the NFTR between two stacked arrangements both of which consist of a repetitive structure of a graphene sheet and a Bi$_2$Se$_3$ slab, as shown in Fig. 16(a).[134] In comparison with the multilayer configuration, the graphene-Bi$_2$Se$_3$ monocell enables wider hybrid polaritons which are taken form due to the coupling of hyperbolic phonon polaritons in Bi$_2$Se$_3$ with surface plasmon polaritons in graphene. Thus, the heat flux of monocell exhibits 1.41 fold and 1.97 fold higher than those of double-cell and ten-cell configurations. In 2022, M.-Q. Yuan et al proposed to utilize a drift-biased graphene grating, which supports nonreciprocal hyperbolic surface plasmon polaritons, to rectify the radiative heat flux between two dielectric particles, as exhibited in Fig. 16(b).[135] Thereafter, C-L Zhou et al studied the NFTR between two parallel monolayer $T_d$-WTe$_2$ sheets with a vacuum gap $d$ [see Fig. 16(c)].[136] Due to the peculiar nature of hyperbolic surface plasmons in $T_d$-WTe$_2$, the heat-transfer coefficient ($h$ = 45.28 kW m$^{-2}$ K$^{-1}$) for $d$ = 10 nm is 3 orders of magnitude larger than that of blackbody limit (6.12 W m$^{-2}$ K$^{-1}$), as depicted in Fig. 16(d), even surpassing that of optimized BP sheets by about 17.6%.[137,138]

Very recently, C.-L. Zhou et al firstly investigated the enhanced NFTR induced by hyperbolic shear polaritons in low-symmetry Bravais crystal $\beta$-Ga$_2$O$_3$.[139] As shown in Fig. 16(e), the NFTR modulator is composed of two $\beta$-Ga$_2$O$_3$ slabs with a vacuum gap $d$ and a twist angle $\Delta\theta$. At $d$ = 10 nm and $\Delta\theta$ = 0° [see Fig. 16(f)], the NFTR modulator produces a heat-transfer coefficient of 50.61 kW m$^{-2}$ K$^{-1}$ that is far larger than those of high-symmetry crystals, e.g., SiC, SiO$_2$, hBN, and the aforementioned $T_d$-WTe$_2$ in Fig. 16(d). Fig. 16(g) further shows the twist-induced NFTR modulation, in which a scaling factor $f$ is defined to describe the magnitude of the off-diagonal component (indicated by $f\varepsilon_{xy}$). At $f$ = 0.0, $\beta$-Ga$_2$O$_3$ becomes a shear-free biaxial crystal that is similar to $\alpha$-MoO$_3$. As the scaling factor $f$ returns to its natural value 1.0, the propagating polaritons become increasingly skewed from the crystal axes. Besides, a twist-induced modulated coefficient $\eta$ is normalized by $h(0°)/h(\Delta\theta)$ for a given $f$ value. It is seen that the twist-induced modulated coefficient is enhanced upon increasing the parameter $f$. As the $f$ returns to its natural value 1.0, the maximal value of $\eta$ increases to 1.27 for $\Delta\theta$ = 90°, which is 1.15 times larger than that in the shear-free scenario. This implies that low-symmetry

Bravais crystal provides a greater potential for noncontact heat dissipation for nanoscale circuits or other devices than traditional

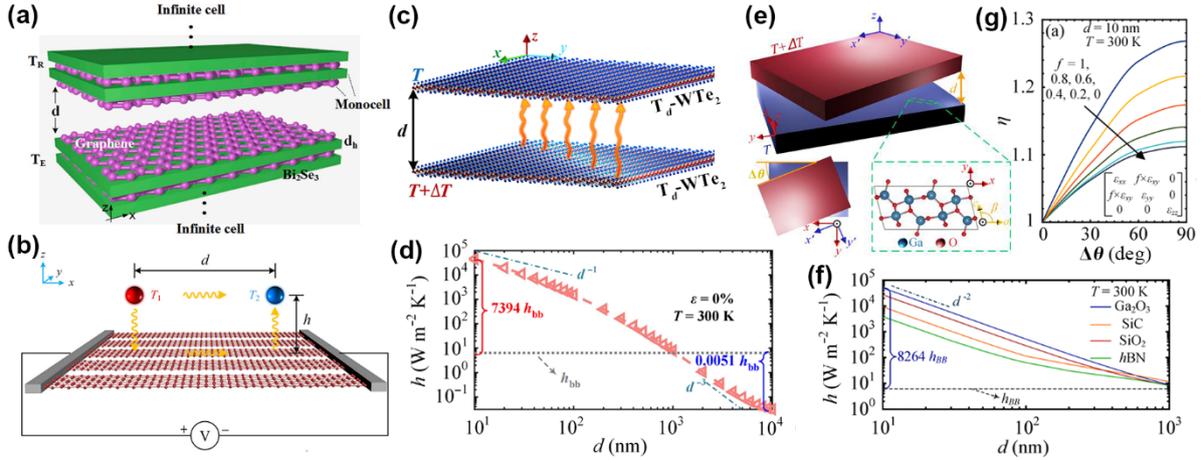

**Fig. 16** (a) Illustration of NFTR between two graphene/Bi$_2$Se$_3$ stacked structures separated by a vacuum gap distance *d*. Reproduced with permission from ref. 134. Copyright 2021, Elsevier. (b) Two particles separated by a distance *d* are placed at a near-field distance of *h* above a graphene grating. A direct-current voltage is applied to both ends of the grating. Reproduced with permission from ref. 135. Copyright 2022, Elsevier. (c) Schematic of NFTR between two monolayer T$_d$-WTe$_2$ sheets separated by a vacuum gap *d*. (d) Heat-transfer coefficient as a function of the vacuum gap *d* for the T$_d$-WTe$_2$ system in (c). The gray dotted line in (d) is the heat-transfer coefficient for the blackbody. (c-d) Reproduced with permission from ref. 136. Copyright 2022, American Physical Society. (e) Setup for NFTR modulator composed of two β-Ga$_2$O$_3$ slabs with the vacuum gap *d* and a twist angle Δϑ. (f) Heat-transfer coefficient of β-Ga$_2$O$_3$ system as a function of *d* at Δϑ = 0° and its comparison with those of SiC, SiO$_2$ and hBN. (g) Twist-induced modulated coefficient *η* at different scaling factors *f* for the magnitude of the off-diagonal component. (e-g) Reproduced with permission from ref. 139.

high-symmetry crystals.

## 5 Summary and outlook

In this work, we have reviewed the recent achievements in 2D natural HMs and their anisotropic properties including anisotropic dispersion relations, hyperbolic phonon, plasmon, exciton, and shear polaritons. The generation mechanism of four polaritons and their discrepancies are compared and discussed in detail. On account of the admirable anisotropic optical properties, many preliminary nanophotonic applications utilizing natural hyperbolicity of 2D crystals have been explored, such as valley quantum interference, mid-infrared polarizer, spontaneous emission enhancement, NFTR, and so forth. However, there are still lots of problems to be solved to attain more in-depth understandings of the properties of 2D natural HMs and to realize their full potential in multifunctional areas. Several potential opportunities and challenges are summarized as follows.

(1) Among four polaritons, experimental researches on hyperbolic phonon polaritons (especially those appear in hBN and α-MoO$_3$)[36,67,70,74,75,77-79,140-149] are much more than the other three polaritons. This could be due to the fact that hyperbolic phonon polaritons are the earliest discovered, and hBN and α-MoO$_3$ are relatively easier to be achieved than the other 2D natural HMs. Besides, 2D natural HMs, which are confirmed by experiments (see Table 1 and Fig. 4), are far less than theoretically predicted ones.[32,33,53,54,150] Furthermore, the dispersion relations for hyperbolic polariton modes are limited by the intrinsic crystal lattice, which are unconducive to the engineering control and frequency tuning of polaritonic modes. To exploit more natural HMs and activate their hyperbolicity in a wider spectral range, it is necessary to attempt more approaches (being crucial to the onset frequency of hyperbolicity and the permittivity[93,151-154]) to modify the electric structures and energy bands of anisotropic layered crystals. The potential approaches include but are not limited to controlling the temperature,[37] straining,[10] doping,[53] photoexcitation,[155] and external stimuli to the dielectric environment of HMs,[49] etc.

(2) As listed in Table 1, the materials of 1T-ZrS$_2$, T$_d$-WTe$_2$, and bismuth support natural hyperbolicity only at low temperatures while the others can hold hyperbolicity at RT. To date, thermal evolution mechanisms of hyperbolic optical properties as well as their discrepancies between different natural HMs have not yet been comprehensively understood.

(3) Based on the density-functional theory, the in-plane component of permittivity for bulk 1T-ZrS$_2$ is negative while that for monolayer 1T-ZrS$_2$ is positive. It means that natural hyperbolicity disappears for monolayer 1T-ZrS$_2$.[156] In contrast, for crystalline T$_d$-WTe$_2$ and BP, both bulk and monolayer counterparts possess their respective hyperbolic regions.[37,84,93,136,152,157] The dependence of natural hyperbolicity on film thickness and their differences between different 2D natural HMs are still bewildering and worth discussing in depth.

(4) Even though metal/HM hybrid architectures have been extensively utilized to modify the optical properties of 2D natural

HMs, they are mainly limited to investigating the coupling and/or hybrids between metal plasmonics and hyperbolic polaritons.[35,158-160] In fact, when the work function difference between the metal and some 2D natural HMs (e.g., tetradymites, WTe$_2$, and α-MoO$_3$)[161-166] is large enough, Schottky barriers could be formed at the metal/HM interface, which can give rise to the energy-band bending and the change of anisotropic permittivity tensors of HMs. Unfortunately, the influence mechanism of Schottky barriers on natural hyperbolicity has not yet been investigated nor reported in the literature.

(5) Hyperbolic shear polaritons are just proposed in this year, and have been only reported for monoclinic crystal β-Ga$_2$O$_3$ so far. The non-diagonalizable dielectric permittivity in low-symmetry crystals plays a crucial role in the realization of hyperbolic shear polaritons. The unique findings related with shear polaritons could be generalized to the other low-symmetry crystals, e.g., monoclinic α-RuCl$_3$ and triclinic ReSe$_2$.[167,168] On the other hand, shear polaritons as well as their coupling with phonon, plasmon, and exciton polaritons may have important applications in the manipulations of phase and directional energy transfer (e.g., ultra-fast asymmetric thermal dissipation in the near field[169] and gate-tunability for on-chip all-optical circuitry[170]), which creates enticing opportunities for more powerful and high-efficiency thermal photons computing and photonic energy-harvesting techniques.

(6) Manipulating natural hyperbolic polaritons not only provides an excellent platform for studying polariton physics, but also inspires people to envision many distinctive application prospects. For instance, deep subdiffractional scales and reconfigurability may endow hyperbolic polaritons with the capability of encoding input optical signal. In the strong coupling regime, hyperbolic polaritons have great potential for non-destructive molecule sensing and point-of-care medical diagnostics.[49]

(7) The uniqueness of 2D natural HMs is their atomic thickness, which is of great significance for highly compact and integrated devices. In future device applications, large-area and high-quality crystallized natural HMs are highly desirable. Currently, high-quality 2D vdW crystals are primarily obtained by mechanical exfoliation which has an apparent drawback of a low yield with the sample size of typically several to tens of micrometers. Thus, it is still quite challenging to directly synthesize 2D natural HMs with a large area and a high-quality crystallization. Besides, the follow-up processing for synthesizing devices may introduce charge carriers and produce defects in natural HMs, which significantly impacts the formation of hyperbolic polaritons. Furthermore, as mentioned above, some 2D materials only support natural hyperbolicity at ultra-low temperatures and/or external stimuli.[22,21,37,49,155] Because of these strict requirements, lots of promising applications (e.g., the aforementioned valley quantum interference, *CP*-breaking induced natural hyperbolicity, enhancement of NFTR based on Bi$_2$Se$_3$, T$_d$-WTe$_2$, β-Ga$_2$O$_3$, etc.) are still in the stage of theoretical predictions. The related experimental observation and even industrial production are highly desirable and require further efforts.

## Conflicts of interest

There are no conflicts to declare.

## Acknowledgements

The authors acknowledge the financial supports by National Natural Science Foundation of China (Grant No. 11804251) and Iran Science Elites Federation (Grant No. M1400138).

## Notes and references